\documentclass[showpacs,amsmath,amssymb,twocolumn,prl,floatfix]{revtex4}
\usepackage[dvips]{graphicx}
\input{epsf}

\begin{document}

\title{Time reversal of Bose-Einstein condensates}

\author{J. Martin, B. Georgeot and D. L. Shepelyansky}

\affiliation{Laboratoire de Physique Th\'eorique,
 Universit\'e de Toulouse III, CNRS, 31062 Toulouse, France}

\date{April 22, 2008. Revised: June 19, 2008.}

\begin{abstract}
Using Gross-Pitaevskii equation,
we study the time reversibility of Bose-Einstein condensates (BEC)
in kicked optical lattices, showing that inside the regime
of quantum chaos the dynamics can be inverted
from explosion to collapse.
The accuracy of time reversal
decreases with the increase of atom interactions inside BEC, until
it is completely lost.  Surprisingly, quantum chaos helps
to restore time reversibility.  These predictions can be tested with
existing experimental setups.
\end{abstract}
\maketitle


In recent years, there has been remarkable progress in the manipulation and
control of the dynamics of BEC in optical lattices
(see e.g. the review \cite{BECrev}).  In such systems,
the velocity spreading
is very small. This allows
to perform very precise investigations of the kicked
rotator, known as a paradigm of quantum and classical chaos
\cite{chirikov}.
As a result, high order quantum resonances were recently observed
experimentally \cite{phillips}, and
various nontrivial effects in the kicked rotator
dynamics were probed \cite{summy,auckland}.  It should be stressed
that in BEC the interactions between atoms are of crucial importance, in
contrast with other implementations of the kicked rotator with cold atoms
\cite{raizen,christensen,darcy,garreau},
where such interactions are negligible. Recently, a method of time reversal
for atomic matter waves has been proposed for the kicked rotator
dynamics of noninteracting atoms \cite{martin}.
This method also allows to realize effective cooling of the atoms by
a few orders of magnitude.
The problem of time reversal of dynamical motion of atoms originates
from the famous dispute between Boltzmann and Loschmidt on the origin
of irreversible statistical behavior in time reversible systems
\cite{loschmidt,boltzmann}.  The results obtained in \cite{martin}
showed that the quantum dynamics of noninteracting atoms can be
reversed in time even if the corresponding classical dynamics
is practically irreversible due to dynamical chaos \cite{sinai,lieberman}.
The investigation of the effect of interactions on time reversibility
is of prime importance, since the original dispute
between Boltzmann and Loschmidt concerned interacting atoms.  In this
paper, we study the effects of interactions between atoms in BEC
on the time reversal accuracy.  We emphasize that interactions bring
new elements in the problem of time reversal compared to the case of
one particle quantum dynamics \cite{martin}, acoustic \cite{finkac}
and electromagnetic waves \cite{finkem}.

To describe the BEC dynamics in a kicked optical lattice we use
the Gross-Pitaevskii equation (GPE) \cite{pitaev}:
\begin{equation}\label{GP}
    i\hbar\frac{\partial}{\partial t}\psi=\left(-\frac{\hbar^2}{2m}\frac{\partial^2}{\partial x^2}
    -g|\psi|^2+k\cos x \; \delta_T(t)\right)\psi
\end{equation}
where the first two terms on the r.h.s.~correspond to the usual GPE and the
last term represents the effect of the optical lattice, with
$\delta_T(t)$ being a periodic delta-function with period $T$.
The interaction between atoms is quantified by the nonlinearity parameter
$g=Ng_{1D}$ where $N$ is the number of atoms in the condensate and
$g_{1D}=-2a_0\hbar\omega_{\perp}$ is the effective 1D coupling constant,
$\omega_{\perp}$ being the radial frequency of the trap and $a_0$ the 3D
scattering length. Here $\psi(x,t)$ is normalized to one. In the
following, we choose units such that $\hbar=m=1$, so that the momentum
of atoms is measured in recoil units; time $t$ is measured in units of $T$.
For $g=0$ Eq.~(\ref{GP}) reduces to the usual kicked rotator model with the
classical chaos parameter $K=kT$ and effective Planck constant $T$
(see e.g.~\cite{martin}).  In this case the time reversal
can be done in the way described in \cite{martin}: the forward
propagation in time is done with $T=4\pi +\epsilon$  while the
backward propagation is performed using
$T\rightarrow T'=4\pi-\epsilon$, with inversion
of the sign of $k$ at the moment of time reversal $t_r$.
This procedure allows
to perform approximate time reversal (ATR) for atoms with small velocities
inside the central recoil cell.
This ATR procedure works quite accurately for noninteracting
atoms ($g=0$), however for real BEC the nonlinear interaction
($g\neq 0$ in (\ref{GP})) may significantly affect the return probability.
Indeed, in the regime of strong interactions and BEC size smaller than
the optical lattice period,
earlier investigations \cite{pikovsky} of soliton dynamics in GPE with $g > 0$
and periodic boundary conditions
showed that with good accuracy a soliton
 moves along chaotic classical trajectories of
the Chirikov standard map \cite{chirikov}.
As a result, even if one performs exact time reversal (ETR),
the presence of very small numerical errors
completely destroys reversibility
due to instability of chaos.
However, here we are mainly interested in the regime typical
of BEC experiments where the BEC size is larger than the lattice period.
Our studies show that in this regime time reversal can be achieved
with good accuracy for moderate strength of interactions.

\begin{figure}
\begin{center}
\includegraphics[width=.95\linewidth]{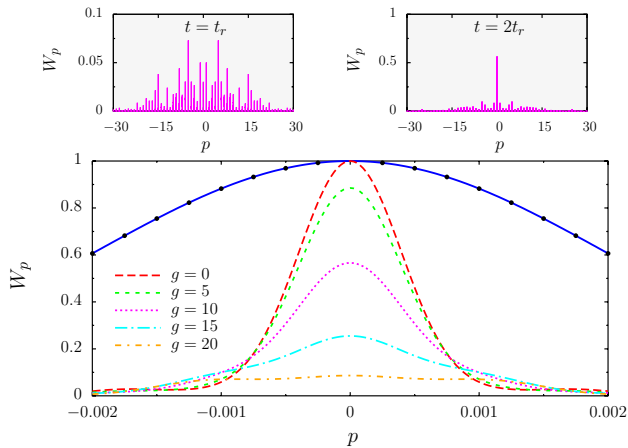}
\end{center}
\vglue -0.50cm \caption{(Color online) Initial (blue/black solid
curve) and final return probability distributions $W_{p}$ obtained
by ATR procedure vs momentum $p$ for various nonlinearities $g$.
Insets : probability distribution $W_p$ at $t=t_r=10$ (left) and
final one at $t=t_f=2t_r$ (right) on a larger scale for $g=10$. All
probability distributions are scaled by their  value at $p=0, t=0$.
Initial state is a Gaussian packet with rms $\sigma=0.002$. Here
$k=4.5$, $T=4 \pi+\epsilon$ for $0\leq t \leq t_r$ and $T'=4
\pi-\epsilon$ for $t_r < t \leq 2t_r $ with $\epsilon=2$ and
$t_r=10$. Black dots correspond to the wave function after ETR
procedure.} \label{fig1}
\end{figure}

To investigate time reversibility for BEC, we numerically simulated
the wave function evolution through (\ref{GP}), using up to
$N_s=2^{23}$ discretized momentum states with $\Delta p=2.5 \times 10^{-5}$,
space discretization $\Delta x= 2\pi/(N_s \Delta p)$,
and with up to $4\times 10^4$ integration time steps between two kicks.
In this way we obtain the probability distribution in momentum space
$W_p(t)=|\langle p |\psi (t)\rangle |^2$.
The results of time reversal performed by ATR procedure are shown
in Fig.1, for an initial Gaussian distribution in momentum with
rms $\sigma=0.002$ in recoil units (as in the experiment of
\cite{phillips}).  After this procedure, the
returned wave packet at $t=2t_r$ becomes clearly squeezed in momentum
compared to the initial one.  In absence of interactions
($g=0$) the maximum of the final returned distribution
$W^f_0=W_{p=0}(t=2t_r)$ is equal
to the maximum of the initial one $W^i_0=W_{p=0}(t=0)$ since ATR is exact
for $p=0$ \cite{martin}.  The increase of nonlinearity parameter $g$ leads
to a reduction of the ratio $W^f_0/W^i_0$ until complete destruction
of reversibility for $g=20$.  However the half-width $p_L$ of the returned
peak is only weakly affected by $g$.  It should be stressed that at
the moment of time reversal $t_r$ the wave packet is completely destroyed
(Fig.1 left inset), and nevertheless the peak is recreated at $t=2t_r$
(Fig.1 right inset).  This process looks similar to the observed
``Bosenova'' explosion induced by the change of sign of the interactions
in BEC \cite{wieman}.  In our case the sign of interactions is unchanged
but the time reversal allows to invert the explosion which happens
during $0\leq t\leq t_r$ into a collapse for $t_r<t \leq 2 t_r$.
We also note that the ETR procedure ($\psi(x)\to\psi^*(x)$ at $t=t_r$)
leads to an almost perfect
time reversal of the wave packet, indicating that exponential instability
is rather weak on the considered time scales.

\begin{figure}
\begin{center}
\includegraphics[width=.95\linewidth]{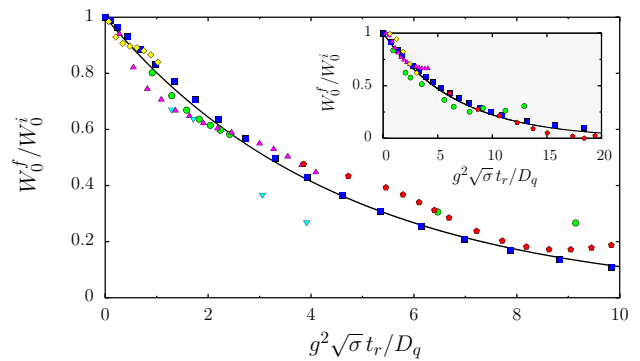}
\end{center}
\vglue -0.50cm \caption{(Color online) Ratio of final to initial
probabilities at $p=0$ as a function of $g^2\sqrt{\sigma}t_r/D_q$.
Symbols mark the numerical data for: $k=4.5$, $\sigma=0.002$,
$t_r=10$ ($0\leq g \leq 20$, blue squares); $k=8$, $g=10$, $t_r=10$
($0.002\leq \sigma\leq 0.2$, green circles); $k=4.5$, $g=10$,
$\sigma=0.002$ ($1 \leq t_r \leq 15$, magenta triangles); $g=10$,
$\sigma=0.002$, $t_r=10$ ($4 \leq k \leq 9$, cyan reversed
triangles); $k=4.5$, $g=10$ ($g=15$ for the inset), $t_r=10$ ($0.002
\leq \sigma \leq 0.026$, red pentagons); $k=4.5$, $g=5$ ($g=15$ for
the inset), $\sigma=0.002$ ($1 \leq t_r \leq 15$, yellow diamonds);
the curve shows the dependence (\ref{gamma}) with $C=0.22$. Inset~:
same ratio but for a change of sign of $g$ during the reversed
evolution, the curve has $C=0.15$.} \label{fig2}
\end{figure}

The dependence of the ratio $W^f_0/W^i_0$ on the system parameters
is shown in Fig.2.  It can be approximately described through
\begin{equation}\label{gamma}
W^f_0/W^i_0 \approx e^{-\Gamma t_r} \; , \;\;\;\Gamma = C g^2 \sqrt{\sigma}/D_q\;,
\end{equation}
where $C$ is a numerical constant and
$D_q$ is the quantum diffusion rate which determines
the localization length $l$ of quantum eigenstates in momentum space
when $g=0$: $l=D_q/2$ with $D_q\approx k^2/2$ (see
exact expression for $D_q$ in \cite{martin}).
Surprisingly enough, the accuracy of time reversal increases
with the increase of chaotic diffusion rate $D_q$ (see Fig.2).
Qualitatively, this is due to the faster spreading of the wave function in
coordinate space, that leads to a decrease of the nonlinear term
$g|\psi(x,t)|^2$ in (\ref{GP}), making the dynamics closer to the linear
case for which $W^f_0 = W^i_0$.
An estimate for $\Gamma$ can be obtained assuming that the nonlinear term
generates additional corrections $k \rightarrow k+a(t)$ in (\ref{GP}),
where $a$ randomly varies in time and
$a \sim g |\psi|^2 \sim g \sigma/\sqrt{l}$.
This relation takes into account the normalization condition
($|\psi|^2/ \sigma \sim 1$) and assumes that inhomogeneities in $|\psi(x)|^2$
are smoothed over $l \sim D_q$ localized chaotic eigenstates.
With these assumptions the
decay rate of $W^f_0/W^i_0$ is given by the Fermi golden rule
$\Gamma \sim |a|^2 \sim g^2\sigma^2/ D_q$. Such a consideration assumes
that $\sigma$ remains constant during the dynamics while in fact
it increases due to diffusion in momentum
that probably leads to the smaller power of
$\sigma$ found numerically (\ref{gamma}).
These estimates qualitatively explain the surprising result
of Fig.~\ref{fig2} which shows that the increase of quantum chaos
diffusion $D_q$ improves the time reversal at fixed $g$.

In addition to the ATR procedure described above,
it is possible to perform additionally the inversion of the sign of
the interactions $g$ at $t=t_r$.
In principle such an inversion of $g$ can be realized in experiments
similar to those of \cite{wieman}.
The numerical data for this case are
shown in the inset of Fig.2.  They can be described by the same
formula as for the case of unchanged $g$, but with a smaller
numerical constant $C$.  The relatively small difference between
the two cases indicates that the main mechanism of time reversal destruction
is related to transitions to other momentum states induced by the
nonlinear interaction.

\begin{figure}
\begin{center}
\includegraphics[width=.95\linewidth]{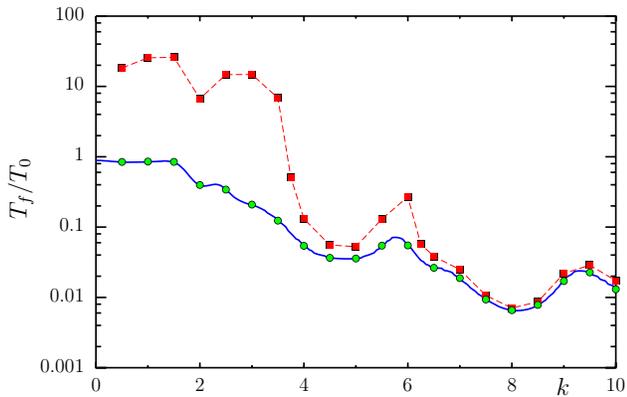}
\end{center}
\vglue -0.50cm \caption{(Color online) Loschmidt cooling of time
reversed BEC atoms characterized by the ratio of final $T_f$ and
initial $T_0$ temperatures (solid curve) as a function of $k$ for an
initial Gaussian momentum distribution with rms $\sigma=0.002$ and
$g=0$ (solid curve), $g=0.5$ (green/gray circles), and $g=10$
(red/black squares). Here $t_r=10$ and $\epsilon=2$.} \label{fig3}
\end{figure}

The time reversal leads to a squeezing of the wave packet in momentum
space near $p=0$, that can be interpreted as an effective Loschmidt cooling
\cite{martin}.  The effect of interactions on this cooling process is
analyzed in Fig.3.  The temperature $T_f$ of the returned atoms
can be defined as the temperature of atoms inside the momentum interval
$[-2p_L,2p_L]$  at $t=2t_r$ (see \cite{martin}).  For $g=0$, the
ratio $T_f/T_0$ of final and initial temperatures
drops significantly with $k$.  At small
$g=0.5$, the ratio remains essentially unchanged for all values of $k$
considered.  In contrast, for stronger nonlinearity ($g=10$), there is no
cooling at low values of $k\leq 3$, but for strong chaos with $k>3$
the cooling reappears and becomes very close to the case
$g=0$ at large $k$ values.  Thus surprisingly strong quantum chaos
enhances the cooling of BEC.  This result is the consequence of
relation (\ref{gamma}), according to which the return probability
becomes larger and larger with the increase of the
quantum diffusion rate $D_q\approx k^2/2$.

The results presented in Figs.1-3 show that time reversal can be performed
for BEC through the ATR procedure even in presence of strong interactions.
The cooling mechanism works in presence of these interactions, if the
quantum chaos is sufficiently strong.

Up to now we discussed the case of spatial width of BEC
much larger than the optical lattice period.  It is also interesting
to analyze the opposite situation where the BEC size becomes
smaller than this period.  In this case we start from the soliton distribution
\begin{equation}\label{soliton}
    \psi(x,t)=\frac{\sqrt{g}}{2}
\frac{\exp\left(ip_0(x-x_0-p_0t/2)+ig^2t/8\right)}
{\cosh\left(\frac{g}{2}(x-x_0-p_0t)\right)} \; .
\end{equation}
For $k=0$ this is the exact solution of Eq.(\ref{GP}), which describes
the propagation of a soliton with constant velocity $p_0$ \cite{refsol}.
At moderate values of $k$ the shape of the soliton is only slightly
perturbed, and its center follows the dynamics described by the
Chirikov standard map \cite{pikovsky}:
$\bar{p}_0=p_0 + k\sin x_0\;;\; \bar{x}_0=x_0+\bar{p}_0 T$, where bars denote
the values of the soliton position and velocity after a kick iteration.
In the chaotic regime with $K>1$ the soliton dynamics becomes truly chaotic.
Indeed, two solitons with slightly different initial velocities
or positions diverge exponentially with time.  As a result of this instability,
even the ETR procedure does not produce an exact return of the soliton to its
initial state, due to the presence of numerical integration errors.
This is illustrated in Fig.4, where the distance in phase space $\delta$
between initial and returned solitons is shown as a function of the
time reversal moment $t_r$.  If the initial position of the soliton
is taken inside the chaotic domain, $\delta$ grows exponentially with
$t_r$ as $\delta \sim \exp (\lambda t_r) $.  The numerical fit
gives a value of $\lambda$ very close to the Kolmogorov-Sinai entropy $h$
of the standard map at the corresponding value of $K$.  For large values of
$t_r$ the values of $\delta$ become so large that the time reversed soliton
is located far away from the initial one, and thus time reversibility
is completely destroyed (Fig.4 left inset).
In contrast, if
the soliton starts in the regular domain, the growth of $\delta$
remains weak during the whole integration time.  In this regime, even
for large values of $t_r$ the values of $\delta$ remain small and the
soliton returns very close to its initial state (Fig.4 right inset).

\begin{figure}
\begin{center}
\includegraphics[width=.95\linewidth]{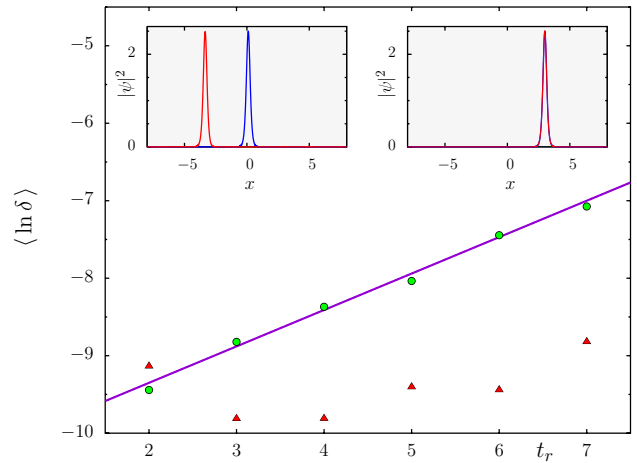}
\end{center}
\vglue -0.50cm \caption{(Color online) Phase space distance
$\delta=\sqrt{\delta x^2+\delta p^2}$ between initial soliton and
its time reversed image obtained by ETR ($\psi(x)\to\psi^*(x)$) vs
$t_r$ for $k=1$, $T=2$ and $g=10$; symbols show numerical data
averaged over 50 trajectories inside the chaotic (green circles) and
the regular (red triangles) domains. The solid line shows the linear
fit $\langle \ln \delta \rangle = 0.47\,t_r-10.29$. The slope is
close to the Kolmogorov-Sinai entropy $h\approx 0.45$ at $K=kT=2$.
Insets~: soliton at initial time $t=0$ (blue/black) and final time
$t=2t_r$ (red/grey) with $t_r=40$ for initial conditions inside the
chaotic domain (left) and regular domain (right, initial and final
solitons are superimposed).} \label{fig4}
\end{figure}

\begin{figure}
\begin{center}
\includegraphics[width=.95\linewidth]{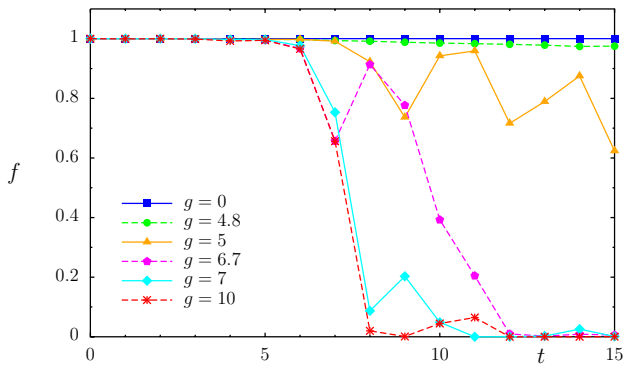}
\end{center}
\vglue -0.50cm \caption{(Color online) Fidelity of two solitons with
close initial conditions vs time $t$ for $k=1$, $T=2$ and different
nonlinearities.
The initial states
are solitons (\ref{soliton})
with $g=10$ and initial conditions
$(x_0,p_0)=(0.5,1.25)$ and $(x'_0,p'_0)=(0.5,1.255)$ inside
the chaotic domain.}
\label{fig5}
\end{figure}

Another way to characterize the stability of nonlinear wave dynamics
described by Eq.(\ref{GP}) is to study the behavior of the fidelity
defined as $f(t)= |\langle \psi_1(t)|\psi_2(t) \rangle|^2$,
where $\psi_1$ and $\psi_2$
are two solitons with slightly different initial conditions.
It is important to note that for $g=0$ the function $f(t)$ is independent
of $t$, thus variation with time appears only due to nonlinear effects.
In a certain sense this quantity can be considered as a generalization
of the usual fidelity \cite{prosen}
discussed for the Schr\"odinger evolution to the
case of nonlinear evolution given by GPE.  The dependence
of $f$ on time is shown in Fig.5 for different values of the nonlinearity
parameter $g$.  For small values of $g$ the function is almost constant on
the considered time interval, while for large values of $g$ it drops quickly
to almost zero after a logarithmically short time scale corresponding to
the separation of the two solitons.  These two qualitatively different
behaviors can be understood as follows.
For relatively weak nonlinearity, $f(t) \sim \exp (-\Gamma t)$
with $\Gamma \propto g^2$ (see Eq.(\ref{gamma})) that corresponds
to the usual Fermi golden rule regime of the fidelity decay
in linear quantum systems with perturbation \cite{prosen}.
For stronger $g$, we enter the regime where nonlinear
wave packets move like chaotic individual particles that leads
to an abrupt drop of fidelity as soon as the separation becomes larger
than the size of wave packets.
In this regime the size of BEC is smaller than the lattice
period and the time reversal is destroyed.
In the opposite limit of large BEC size shown in Fig.~\ref{fig1}
the  time reversal can be maintained at moderate values of $g$.
The transition between these two regimes
is rather nontrivial and requires further investigations.

The results of Figs.4-5 show that the soliton dynamics described
by GPE (\ref{GP}) is truly chaotic.
This leads to the destruction of time reversibility
induced by exponential growth of small perturbations.
However, the real BEC is a quantum object
with a mass proportional to the number of atoms $N$.  Thus it has
effective $\hbar_{\mathrm{eff}}\propto 1/\sqrt{N} $ and since the
Ehrenfest time $t_E$ for chaotic dynamics depends only logarithmically
on $\hbar$ \cite{chirikov,prosen} this time  remains rather short
$t_E \sim |\ln \hbar_{\mathrm{eff}}|/h \sim (\ln N)/2h$. Thus for
conditions of Fig.4 and $N=10^5$ we have $t_E \approx 13$ and
on larger times quantum BEC should have no exponential instability
of motion.  Hence, in the time reversal procedure
the real quantum BEC remains stable and reversible
contrary to the BEC described by GPE.
We think that the resolution of this paradox relies on the
absence of second quantization in GPE that makes
the soliton dynamics essentially classical.

The dimensionless nonlinear parameter is $g=-Na_0\lambda/2\pi l_{\perp}^2$,
where $l_{\perp}=\sqrt{\hbar/m\omega_{\perp}}$ and $2\lambda$ is the optical
lattice period.  Values of $g \approx 0.1$ have been
achieved at radial frequency $800$Hz and $N\approx 5000
\approx N_{max}= l_{\perp}/|a_0|$\cite{hulet}.
This value can be further increased
up to $g\approx 10$ by increasing $\omega_{\perp}$ (e.g. $20$ kHz
\cite{hinds}) and increasing $\lambda$ using a
CO$_2$-laser \cite{hansch} or tilted laser beams.
Thus experimental setups similar to
\cite{phillips,summy,auckland,hulet} can test the fundamental
question of BEC time reversal discussed here.

We thank CalMiP for access to their supercomputers and
the French ANR (project INFOSYSQQ) and the EC project EuroSQIP for support.

\vglue -0.50cm

\end{document}